\documentclass{elsart}

\usepackage{psfig}

\usepackage{natbib}

\begin{document}

\runauthor{Marshall}

% -------------------------------------------------------------------------

\begin{frontmatter}

\title{Early Results from the Chandra X-ray Observatory}

\author[MIT]{H.L. Marshall}
\address[MIT]{Center for Space Research,
Massachusetts Institute of Technology, Cambridge, MA 02139, USA}

\begin{abstract}
We present some early results on AGN from the Chandra
X-ray Observatory, highlighting high resolution
spectroscopy using the High Energy Transmission Grating Spectrometer
(HETGS).  The quasar PKS~0637--752 was found to have a very bright
X-ray jet whose shape is remarkably similar to that of the radio jet on
a size scale of 100 kpc, but the X-ray emission is still inexplicably 
bright.  Two BL Lac objects, PKS~2155--304 and Mrk~421, observed
with the spectrometer were found to have no strong absorption or
emission features.  Other radio loud AGN observed with the HETGS show
simple power law spectra without obvious features.
\end{abstract}

\begin{keyword}
galaxies: active; quasars: general; X-rays: galaxies
\end{keyword}

\end{frontmatter}

% -------------------------------------------------------------------------

\section{Introduction}

The Chandra X-ray Observatory (CXO, or Chandra) was launched in July,
1999 into a high Earth orbit.
Although there has been some degradation of the spectral resolution
of the Advanced CCD Imaging Spectrometer (ACIS),
all instruments are performing well.
Here I will focus on data from ACIS and the High Energy Transmission
Grating Spectrometer (HETGS), in order to
demonstrate the overall performance of the telescope and some
of the ways that Chandra will contribute to our understanding
of active galactic nuclei (AGN).

\begin{figure}[htb]
\centerline{\psfig{figure=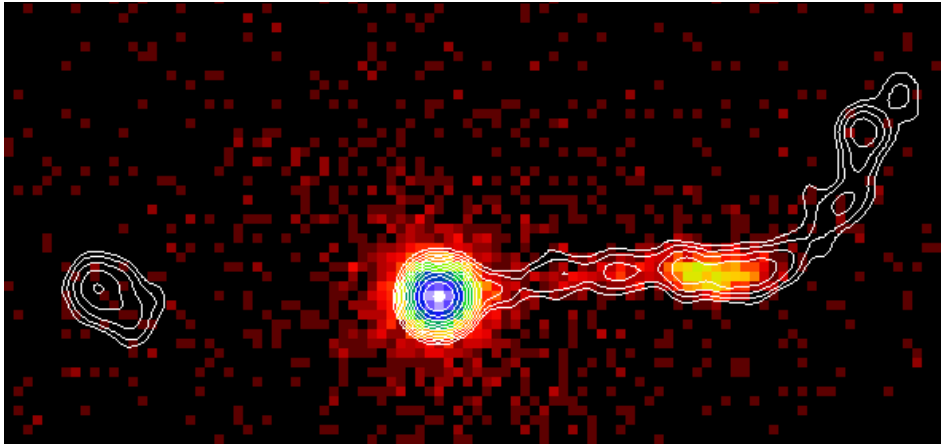}}
\caption{Chandra ACIS image of the focussing target, PKS~0637-752.
Radio flux contours from the Australia Telescope Compact Array are overlaid.
There are three faint sources detected in the Hubble Telescope images
that fall within the brighter parts of the jet, which was not detected
in X-rays or optical light before the launch of Chandra.  The X-ray
brightness of the jet is still unexplained by conventional models.}
\label{fig:jet}
\end{figure}

\section{Chandra Observations of AGN}

The first target observed with Chandra was the quasar
PKS~0637--752 (figure~\ref{fig:jet}).  It was rather surprising
to find an X-ray
jet extending 7-12" from the quasar core because
the target had been chosen for focus measurements.
The core point response function is small enough ($\sim$ 0.42",
half power radius), however, that
the jet had no effect on determining the best detector focus
position.  New radio observations showed that the radio and
X-ray jets were remarkably similar out to 12" where there
is a significant bend in the radio emission (figure~\ref{fig:jet}).
The optical fluxes from the Hubble data are so low that
models of the jet X-ray emission are difficult to constrain. 
\cite{SM00}.

\begin{figure}
\centerline{\psfig{figure=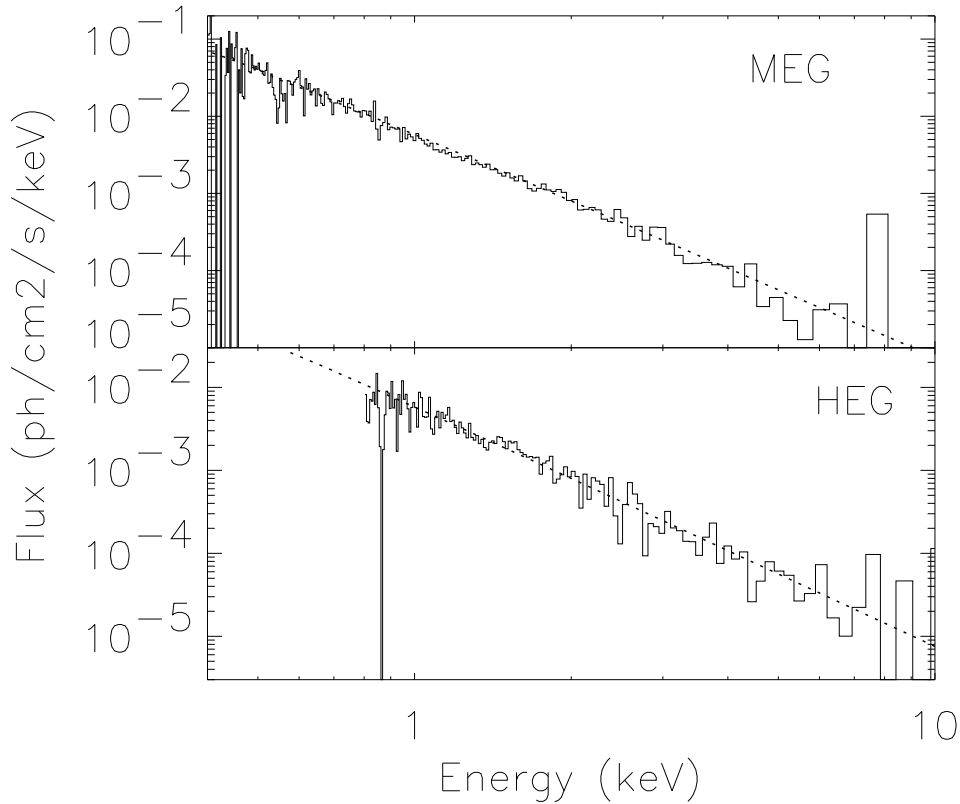}}
\caption{The MEG (top) and HEG (bottom)
spectra of Mrk~421 with the Chandra HETGS.
The spectra are overplotted with well-fitting
power law models with $\alpha = 1.9$.
The MEG and HEG spectra are consistent within
10-20\% systematic uncertainties.  There are no significant
features in the spectrum except for an instrumental residual
near the O-K edge at about 0.54 keV.}
\label{fig:mk421}
\end{figure}

The HETGS flight calibration program included observations of the
late type star Capella
in order to verify the spectral resolution by using emission lines,
and an observation of the BL Lac object Mrk~421 was included 
in order to verify the effective area for point sources.
An observation of 3C~273 was scheduled for January, 2000, which
will be used for cross-calibration between Chandra spectrometers
and other X-ray telescopes, including BeppoSAX and ASCA.
The Capella observation \cite{C00} shows that the
spectral resolution meets the pre-flight expectations.
The calibration spectra of Mrk~421 are shown in Figure~\ref{fig:mk421}.
The spectra from the MEG and the HEG are consistent with each
other to within the 10-20\% systematic uncertainties.  The
spectra are well fitted by a simple power law with
$\alpha = 1.9$ ($f_{\nu} \propto \nu^{-\alpha}$),
and the only deviation from this fit appears
near the O-K edge, which will be removed as the
detector calibration is refined.

\begin{figure}
\centerline{\psfig{figure=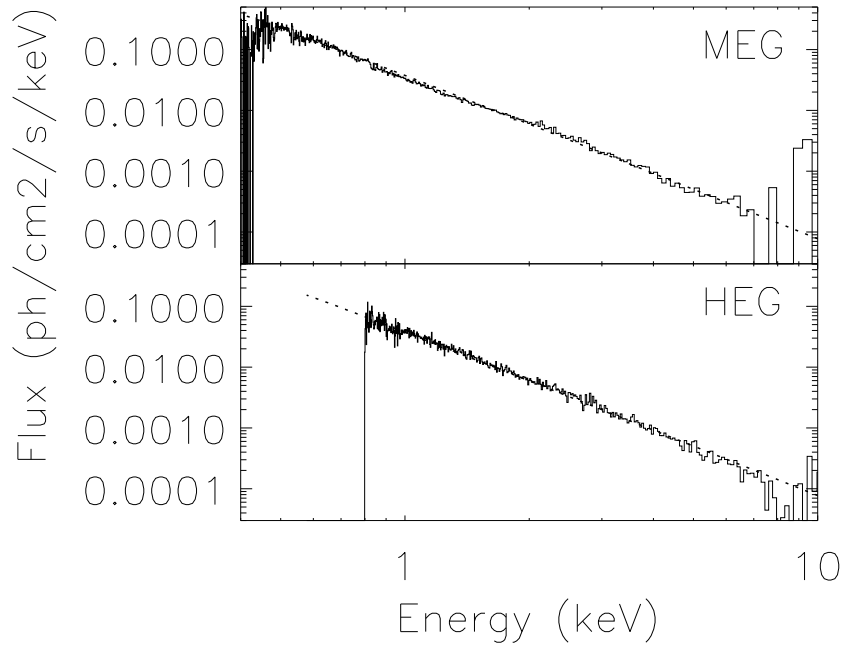}}
\caption{The spectrum of PKS~2155--304 with the Chandra HETGS.
As with Mrk~421, the MEG and HEG spectra are consistent to within
the 10-20\% systematic uncertainties; the spectra are fitted very
well by a simple power law $\alpha = 1.7$.}
\label{fig:pks2155}
\end{figure}

The BL Lac object PKS~2155--304 was observed as part of the
HETGS guaranteed time observation program.  Spectra are
shown in Figure~\ref{fig:pks2155}.  Again, the HEG and MEG
spectra are consistent to within the 10-20\% systematic unceratainties
and are well fitted by a simple, pure power law model.
The spectral index is $1.70 +/- 0.02$ and there are no
significant absorption features.  A feature such as the one
found previously \cite{CK84} would have been detected
easily in the MEG portion of the Chandra HETGS spectrum, so
we conclude that it must be variable.

Other AGN observed in the early phase of Chandra HETGS
guaranteed time observations include NGC~1275 and PKS~2149-305.
Preliminary analysis indicates that these HETGS spectra
are all well fitted by simple power law models with
absorption by neutral interstellar material.  The spectral
indices are somewhat smaller: 0.8 and 0.2, respectively.
No Fe-K$\alpha$ lines are detected in any of their spectra.

\section{Acknowledgements}

I thank the Principal Investigator of the HETGS,
Prof. Claude R. Canizares
for his support and the rest of the HETGS team for their
contributions to the development of the HETGS.
This work was supported in part by the contract SAO SV1-61010.

% -------------------------------------------------------------------------

% -------------------------------------------------------------------------

\end{document}